\documentclass[12pt, a4paper]{article}
\usepackage{fullpage}
\usepackage{amsmath}
\usepackage{verbatim}
\usepackage[dvips]{graphicx}
\usepackage{algorithm}
\usepackage{cite}
\usepackage{setspace}

\newtheorem{definition}{Definition}

\graphicspath{{figures/}}

\title{A polynomial graph extension procedure for improving graph isomorphism algorithms}
\author{Daniel Cosmin Porumbel}

\begin{document}
\maketitle


\begin{abstract}
We present in this short note a polynomial graph extension procedure that can be used to improve any graph isomorphism algorithm. This construction propagates new constraints from the isomorphism constraints of the input graphs (denoted by $G(V,E)$ and $G'(V',E')$). Thus, information from the edge structures of $G$ and $G'$ is "hashed" into the weighted edges of the extended graphs. A bijective mapping is an isomorphism of the initial graphs if and only if it is an isomorphism of the extended graphs. As such, the construction enables the identification of pair of vertices $i\in V$ and $i'\in V'$ that can not be mapped by any isomorphism $h^*:V \to V'$ (e.g.\ if the extended edges of $i$ and $i'$ are different). A forbidding matrix $F$, that encodes all pairs of incompatible mappings $(i,i')$, is constructed in order to be used by a different algorithm. 
Moreover, tests on numerous graph classes show that the matrix $F$ might leave only one compatible element for each $i \in V$.

\end{abstract}

\section{Introduction and Notations}

In theoretical computer science, GI is one of the only $NP$ problems that is not known to be either in $P$ or $NP-P$ (we assume $P\neq NP$) and a lot of effort has been done to classify it. Proofs of  polynomial time algorithms are available for many graph classes~\cite{Luks:01,bge,bab:02}, but, however, all existing algorithms are still exponential for some well-known families of difficult graphs, e.g.\ regular graph isomorphism is GI-complete~\cite{rus,reg} (if regular graphs can be tested for isomorphism in polynomial time, then so can be any two graphs). 

We denote the adjacency matrices of $G$ and $G'$ by M and M'. The number of vertices (denoted by $|G|$ or $|V|$) is commonly referred to as the graph order. A mapping between $G$ and $G'$ is represented by a bijective function on the vertex set $h:V\rightarrow V'$. We say that $h^*$ is an isomorphism if and only if $(i,j)\in E \Leftrightarrow (h^*(i),h^*(j))\in E'$ and the \emph{graph isomorphism (GI) problem} is to decide whether or not such an isomorphism exists.

A critical problem of all tested algorithms appears in the following situation: if there is no edge between vertex $i$ and $j$ in G (i.e. $M_{i,j}=0$) and no edge between $h(i)$ and $h(j)$ in $G'$, than the assignment $(i,j)\stackrel{h}{\to}(h(i),h(j))$ is not seen as a conflict---there is no mechanism to directly detect whether ($i,j$) and $(h(i),h(j))$ are indeed compatible or not. But, by exploiting the structure of the graph, one can find many conditions in which $(i,j)$ and $(h(i),h(j))$ are incompatible even if they are both disconnected (e.g. by checking the shortest path between them).

\section{Polynomial graph extension}

We define the $|V|\times|V|$ matrix $N^{\alpha}$, in which the element $N^{\alpha}_{i,j}$ is the number of paths of length $\alpha$ (i.e. with $\alpha$ edges) from $i$ to $j$. Obviously $N^1=M$, and we now show that $N^{\alpha+1}$ can be computed in polynomial time from $M$ and $N^\alpha$ using the following algorithm:

\begin{algorithm}[ht]
 {\fontfamily{pcr}\selectfont
   \begin{list}{}{\leftmargin 1em \topsep 0pt \parsep 0pt \itemsep 0pt}\parskip 0pt 
      \item \textbf{Input}: $M$ and $N^\alpha$
      \item \textbf{Result}: $N^{\alpha+1}$
      \begin{enumerate}
      \item Set all elements of $N^{\alpha+1}$ to $0$
      \item \textbf{For} $i=1$ \textbf{to} $|V|$
            \begin{description}
            \item[] \textbf{For} $j=i+1$ \textbf{to} $|V|$
            			\begin{description}
            			\item[] \textbf{If} $M[i,j]=1$ \textbf{then}
            						\begin{description}
            						\item[] \textbf{For} $k=1$ \textbf{to} $|V|$
            								\begin{itemize}
            								\item $N^{\alpha+1}[i,k]=N^{\alpha+1}[i,k]+N^{\alpha}[j,k]$
            								\item $N^{\alpha+1}[j,k]=N^{\alpha+1}[j,k]+N^{\alpha}[i,k]$
            								\item $N^{\alpha+1}[k,j]=N^{\alpha+1}[j,k]$ and $N^{\alpha+1}[k,i]=N^{\alpha+1}[i,k]$
            								\end{itemize}
            						\end{description}
            			\end{description}
            \end{description}
      \end{enumerate}
   \end{list}
 }
\caption{Graph extension in polynomial time \label{alg:gex}}
\end{algorithm}

The extended graph is straightforwardly defined as the weighted graph with vertex set $V$ and weighted edges $E^\alpha$ such that if $N^{\alpha}_{ij}\neq 0$, then $\{i,j,N^{\alpha}_{ij}\} \in E$. Two graphs $G$ and $G'$ are isomorphic if and only if their extended graphs are isomorphic---because the same extending operations are applied in the same manner for any two isomorphic vertices $i$ and $h^*(i)$. 

An important advantage of constructing all matrices $N^1, N^2, \dots N^\alpha$ is the early detection of incompatible (forbidden) assignments, i.e. vertices $(i\in V,i'\in V')$ that can never be mapped by an isomorphism. 

\begin{definition}(Compatible assignment) Vertices $i\in V$ and $i' \in V'$ are \emph{compatible} if and only if: (i) $N^\alpha_{i,i}=N'^\alpha_{i',i'}$ and (ii) all the values from line $i$ of $N^\alpha$ can be found in line $i'$ of $N'^\alpha$ and vice versa.
\end{definition}

Indeed, if $h^*$ is an isomorphism, then all assignment $(i\rightarrow h^*(i))$ are compatible; each element $(i,j)$ of line $i$ of $N^\alpha$, can also be found at position $(h^*(i),h^*(j))$ in line $h^*(j)$ of $N'^\alpha$. Therefore, any GI algorithm should never map two incompatible (forbidden) vertices. We introduce a matrix $F$ encoding forbidden mappings, i.e. if $F_{i,i'}=1$, $i$ is never mapped to $i'$. This matrix is empty at start (all elements are 0), and the extension algorithm gradually fills its elements while constructing the matrices $N^1, N^2, \dots N^\alpha$.

The matrices $N^1$, $N^2$, $N^3$, $\dots$ are very rich in information that is implicitly checked via the matrix $F$. Each edge value from the extended graph is in fact a a hash function of some larger structures in the initial graph. Indeed, the fact that an assignment $i \rightarrow i'$ is not forbidden (i.e.\ $F_{ii'}=0$) implies numerous hidden conditions: $i$ and $i'$ need to have the same degree (otherwise $N^2_{i,i}\neq N'^2_{i',i'}$), they need to be part in the same number of triangles (otherwise, $N^3_{i,i}!=N'^3_{i',i'}$), they need to have the same number of 2-step neighbors, etc. Many other such theoretical conditions can be derived and proved, but the goal of this specific paper is only to present a very practical, high-speed algorithm; such theoretical conditions are investigated in greater detail in a completely different theoretical study.

Finally, we note that our practical C++ implementation uses unsigned long integer variables encoded on 64 bits. However, for large values of $\alpha$, $N^\alpha_{i,j}$ can exceed $2^{32}-1$; therefore, we consider all addition operations Modulo $2^{32}$ (in our C++ version, the variables are encoded so that $2^{32}-1+1=0$). This observation does not change the fact that $f^\alpha(h^*)=0$ when $h^*$ is an isomorphism, because if $N^\alpha_{i,j}= N'^\alpha_{h(i),h(j)}$, then $N^\alpha_{i,j}= N'^\alpha_{h(i),h(j)}$ ($Modulo$ $2^{32}$). However, it is still theoretically possible to have the Modulo equality without the non-Modulo equality.

\section{Conclusion}\label{sec:al}

We implemented several algorithms, both exact and heuristics using this property (especially the matrix $F$).
Generally speaking, such an algorithm consists in two stages: (i) the graph extension (ii)the effective algorithm that can be quite naive. The first stage builds the information-rich adjacency matrix $N^\alpha$ and it also provides a matrix $F$ of forbidden vertex assignments.

Numerous tests of such an algorithm with very large graphs show worst-case behavior of polynomial time. Only the strongly regular graphs can show more difficulties, but we tested only several strongly regular graphs with up to 275 vertices and the behavior seems similar. Larger strongly regular graphs are not classified, and since there is no practical algorithm to generate them, we restricted to examples publicly available on the Internet---the McLaughlin Graph with 275 vertices.

The \emph{extending} procedure provides additional evidence that the graph isomorphism can be (at least in practice) solved in polynomial time for almost all graph types we know.

\bibliographystyle{plain}

\begin{thebibliography}{1}

\bibitem{bab:02}
L.~Babai, D.Y. Grigoryev, and D.~M. Mount.
\newblock Isomorphism of graphs with bounded eigenvalue multiplicity.
\newblock In {\em Fourteenth Annual ACM Symposium on Theory of Computing},
  pages 310--7324, 1982.

\bibitem{reg}
K.S. Booth.
\newblock Isomorphism testing for graphs, semigroups, and finite automata are
  polynomially equivalent problems.
\newblock {\em SIAM Journal of Computing}, 7(3):273--279, 1978.

\bibitem{bge}
I.~S. Filotti and J.~N. Mayer.
\newblock A polynomial-time algorithm for determining the isomorphism of graphs
  of fixed genus.
\newblock In {\em STOC '80: Proceedings of the twelfth annual ACM symposium on
  Theory of computing}, pages 236--243. ACM, 1980.

\bibitem{Luks:01}
E.~M. Luks.
\newblock Isomorphism of graphs of bounded valence can be tested in polynomial
  time.
\newblock {\em Journal of computer and system sciences}, 25(1), 1982.

\bibitem{rus}
V.~N. Zemlyachenko, N.~M. Korneenko, and R.~I. Tyshkevich.
\newblock {Graph isomorphism problem}.
\newblock {\em Journal of Mathematical Sciences}, 29(4):1426--1481, 1985.

\end{thebibliography}

\end{document}